
\overfullrule=0pt
\magnification=\magstep1
\font\twelvebf=cmbx12
\nopagenumbers
\line{\hfil CU-TP-540}
\line{\hfil FERMILAB-Pub-91/326-A\&T}
\vglue .5in
\centerline {\twelvebf A Classical Instability of
Reissner-Nordstr\"om Solutions}
\vskip .1in
\centerline {\twelvebf and the Fate of Magnetically
Charged Black Holes}
\vskip .3in
\centerline{\it Kimyeong Lee,$^{a*}$  V.P.Nair$^a$, and
Erick J. Weinberg$^{a,b}$}

\vskip .1in
\centerline { ${}^a$Physics Department, Columbia University}
\centerline {New York, New York 10027}
\vskip .1in
\centerline{${}^b$Theory Group and NASA/Fermilab Astrophysics Center}
\centerline{Fermi National Accelerator Laboratory}
\centerline{P.O.Box 500, Batavia, Illinois 60510.}
\vskip .8in
\baselineskip=16pt
\overfullrule=0pt
\centerline {\bf Abstract}
\vskip .1in
      Working in the context of spontaneously broken gauge
theories, we show that the magnetically charged
Reissner-Nordstr\"om solution develops a classical instability if
the horizon is sufficiently small.  This instability has
significant implications for the evolution of a magnetically
charged black hole.  In particular, it leads to the possibility
that such a hole could evaporate completely, leaving in its place
a nonsingular magnetic monopole.

\noindent\footnote{}{This work was supported in part by the US Department of
Energy (VPN and EJW), by NASA (EJW) under grant NAGW-2381 and by an NSF
Presidential Young Investigator Award (KL).}
\noindent\footnote{}{* Alfred P. Sloan Fellow.}
\vfill\eject

\magnification=\magstep1
\baselineskip=18pt

\catcode`@=11
\def\@versim#1#2{\lower.7\p@\vbox{\baselineskip\z@skip\lineskip-.5\p@
    \ialign{$\m@th#1\hfil##\hfil$\crcr#2\crcr\sim\crcr}}}
\def\simge{\mathrel{\mathpalette\@versim>}} %
\def\simle{\mathrel{\mathpalette\@versim<}} %
\catcode`@=12 

\def\pr#1#2#3#4{Phys. Rev. D{\bf #1}, #2 (19#3#4)}
\def\prl#1#2#3#4{Phys. Rev. Lett. {\bf #1}, #2 (19#3#4)}

\def\np#1#2#3#4{Nucl. Phys. {\bf B#1}, #2 (19#3#4)}
\def\pl#1#2#3#4{Phys. Lett. {\bf #1B}, #2 (19#3#4)}

\def\e{{\bf \hat e}}
\def\a{{\bf  a}}
\def\A{{\bf  A}}
\def\F{{\bf  F}}
\def\Pphi{{\bf  \Phi}}

\def\L{{\cal L}_{ Matter}}

      The Reissner-Nordstr\"om solution to the coupled Einstein-Maxwell
equations describes a spherically symmetric black hole endowed with
electric or magnetic charge.  Although the solution makes
mathematical sense in a theory involving only gravity and the
electromagnetic field, its physical motivation is somewhat tenuous
unless the theory also contains particles carrying such charges.
This remark has little consequence in the case of electric charge,
since one only need add a field whose elementary particles are
electrically charged.   The addition of magnetic charge to the
theory is less trivial, but can be accomplished by enlarging the
structure of the theory so that electromagnetism emerges as the
unbroken subgroup of a spontaneously broken gauge theory in which
magnetic monopoles arise as topologically nontrivial classical
solution~[1].  The incorporation of these additional fields
into the Reissner-Nordstr\"om solution is rather
straightforward~[2], and changes neither the metric nor the
magnetic field. However, as we will show in this letter, the
presence of these fields can render this solution unstable.  This
instability arises at the level of the classical field equations
and does not depend on any quantum mechanical process.   It has
important implications for the ultimate fate of magnetically
charged black holes.

      The magnetically charged Reissner-Nordstr\"om solution to the
Maxwell-Einstein equations has a radial magnetic field with
magnitude  $Q_M /r^2$ and a metric which may be written as
$$ ds^2 = B dt^2 - A dr^2 - r^2 d\theta^2
    - r^2\sin^2\!\theta d\phi^2
     \eqno(1) $$
where
$$ B  = A^{-1}  =  1 - {2MG \over r} + {4\pi G Q_M^2 \over r^2}
     \equiv B_{RN}
     \eqno(2) $$
There is a physical singularity at $r=0$ which is hidden within a
horizon at
$$ r_H = MG + \sqrt{M^2G^2 -4\pi G Q^2_M}
    \eqno(3)$$
provided that the mass $M$ is greater than
$$    M_{crit} = \sqrt{4\pi} |Q_M| M_P
     \eqno(4)$$
where the Planck mass $M_P = G^{-1/2}$.  If $|Q_M| \gg
1$ (as will be the case for the weak gauge coupling we will
assume) the horizon of the critical Reissner-Nordstr\"om black
hole is at $r_H \gg M_P^{-1}$, and is thus in a region where
quantum gravity effects can be neglected.

      This solution is readily incorporated into a theory
possessing classical magnetic mono\-pole solutions.
For definiteness we consider an $SU(2)$ gauge theory which
is spontaneously broken to $U(1)$ by the vacuum expectation value
of a triplet Higgs field $\Pphi$.   The action is
$$ S = \int d^4x \sqrt{-g} \left[ -{1\over 16\pi G}R + \L \right]
     \eqno(5) $$
where
$$  \L = -{1\over 4}\F_{\mu\nu}\cdot \F^{\mu\nu}
         +{1\over 2} D_\mu\Pphi \cdot D^\mu\Pphi
         - V(|\Pphi|)
     \eqno(6) $$
$$ \F_{\mu\nu} = \partial_\mu \A_\nu -\partial_\nu \A_\mu
          - e \A_\mu \times \A_\nu
     \eqno(7) $$
$$ D_\mu \Pphi = \partial_\mu \Pphi -e \A_\mu \times \Pphi
     \eqno(8) $$
and vector notation refers to the internal $SU(2)$ indices.  The
potential $V(|\Pphi|)$ is assumed to have a minimum at $|\Pphi| =
v $; to avoid a cosmological constant, $V$ must vanish at this
minimum.   This theory contains nonsingular monopoles with
magnetic charge $Q_M = 1/e$ and mass $M_{mon} \sim 4\pi v/e$,
provided that $v\simle M_P$.  (For $v$ larger than this, the
would-be monopoles are so massive that they become black holes
themselves~[3,4].)

     The metric for the Reissner-Nordstr\"om solutions of this
theory is precisely the same as that given above for the Maxwell
theory.  For vanishing electric charge and magnetic charge $Q_M =
n/e$, the matter fields are, up to a possible gauge
transformation,
$$ \Pphi = v\, \e(\theta,\phi)
     \eqno(9) $$
$$ \A_\mu = {1\over e} \e \times \partial_\mu \e
     \eqno(10) $$
where $\e$ is a unit vector with winding number $n$; a convenient
choice, which we adopt henceforth, is $\e = (\sin\theta \cos n\phi,
\sin\theta \sin n\phi, \cos\theta)$.  These imply that
$$ \F_{\theta\phi} = - \F_{\phi\theta}
     =  {n \over e}\sin\theta \, \e
     \eqno(11) $$
This lies entirely within the electromagnetic $U(1)$ subgroup defined
by the Higgs field and precisely reproduces the radial magnetic field
of the Maxwell-Einstein theory.   All other components of the field
strength, as well as all of  the covariant derivatives of $\Pphi$,
vanish.

     We now investigate the stability of these solutions, beginning with
the case of perturbations about the solution with unit
magnetic charge.  The problem can be simplified by considering only
spherically symmetric configurations; this turns out to be
sufficient to demonstrate instability.   For such configurations,
the metric can be written in the form of Eq.(1), with $B$ and $A$
being functions only of $r$ and $t$. By an appropriate gauge choice the
matter fields can be brought to the form
$$ \Pphi = v\, \e(\theta,\phi) \,h(r,t)
     \eqno(12) $$
$$  \A_i = {1\over e} \,\e \times \partial_i \e\, (1-u(r,t))
     \eqno(13) $$
where $A_0=0$ because we are interested in electrically neutral
solutions.
(With our choice for $\e$, this reduces to the standard ansatz
in flat space.)  Substitution of this into the matter
Lagrangian gives
$$  \L =  {1\over B} \left[ {{\dot u}^2 \over e^2r^2}
        + {1\over 2}v^2 {\dot h}^2 \right]
      -{1\over A} \left[ {{u'}^2 \over e^2r^2}
        + {1\over 2}v^2 {h'}^2 \right]
       -  {(u^2-1)^2\over 2e^2r^4} - {u^2h^2v^2\over r^2}
            -V(h)
     \eqno(14) $$
with overdots and primes referring to derivatives with respect to
$t$ and $r$, respectively. This leads to the equations
$$  {1\over \sqrt{AB}}{\partial\over \partial t} \left( \sqrt{AB}
         \dot h\over B\right)
   -{1\over r^2\sqrt{AB}}{\partial\over \partial r}
      \left(r^2\sqrt{AB}
         h^\prime\over A\right)
      = -{2hu^2\over r^2} -  {1\over v^2}{dV\over dh}
       \eqno(15) $$
and
$$ {1\over \sqrt{AB}}{\partial\over \partial t} \left( \sqrt{AB}
         \dot u\over B\right)
   -{1\over \sqrt{AB}}{\partial\over \partial r} \left( \sqrt{AB}
         u^\prime\over A\right)
     = -{u(u^2-1)\over r^2} - e^2uh^2v^2
     \eqno(16) $$
for the matter fields, as well as equations, whose explicit form
we do not need, for the metric coefficients $A$ and $B$.  To
consider fluctuations about the Reissner-Nordstr\"om solution we
only need keep terms linear in $u$, $h-1$, $\delta B \equiv
B-B_{RN}$, and $\delta A \equiv A- B^{-1}_{RN}$.  Remarkably, the
coupled equations separate.   The equations for the metric
components contain neither $u$ nor $h-1$, and thus cannot lead to
unstable modes (otherwise there would be an instability in the
pure Maxwell-Einstein case).  The perturbation of the scalar
field enters only in the linearization of Eq.~(15), and can be
shown not to lead to instability.  The remaining fluctuation,
$u$, is determined by the linearized version of Eq.~(16).  If we
define a variable $x$  by
$$ {dx \over dr} = {1 \over B_{RN}(r) }
     \eqno(17)$$
so that $x$ ranges from $-\infty$ to $\infty$ as one goes from
the horizon to spatial infinity, then the equation for $u$ may be
written as
$$ 0= \ddot u - {d^2u\over dx^2} +U(x) u
     \eqno(18)$$
where
$$  U(x) =  B_{RN}(r) { (e^2v^2r^2 - 1)\over r^2}
    \eqno(19)  $$
and $r$ is understood to be a function of $x$ determined by
Eq.~(17). Instability occurs if there are solutions of the form
$u(r,t) = f(r) e^{\omega t}$ with real $\omega$.  Substitution of
this gives a one-dimensional Schroedinger equation for a particle
moving under the influence of the potential $U(x)$.  The unstable
mode exists if this potential has a bound state.
Since $U(x)$ goes to the positive value $e^2v^2$ at $x=\infty$, although
it goes to zero at $x=-\infty$, it is not entirely trivial to see for
what range of parameters we have a bound state.
One can show that a bound state exists if
$r_H < c(ev)^{-1}$ where
$c$ is somewhat less than one, or,
equivalently for
$$   M <  {cM_P^2 \over 2ev} + {2\pi v\over ce}
      \eqno(20)$$
As $M\rightarrow M_{crit}$, $c$ approaches unity.
For $M\gg M_{crit}$, we can
bound $c$ by a variational calculation.
Using the variational ansatz $u=\sqrt{r-r_H}\exp (-\lambda
(r-r_H)/2)$, we find $c > 0.32$.

    The physical basis for this instability is easily understood.
The classical monopole solution has a core of radius $\sim
(ev)^{-1}$, inside of which the Higgs field deviates from its
vacuum value and the massive components of the gauge
field are nonzero.  The effect of this core is to
remove the singularity in the energy density which would arise
from a point magnetic charge.  Its radius is determined by the
balancing of the energy needed to produce the nontrivial
matter fields against the energy cost of extending the Coulomb
magnetic field further inward.

      Similar considerations can be applied to solutions with
horizons.  Here, however, we should only consider the region
outside the horizon since singularities are allowed, and even
expected, inside the horizon.  Looking at the case of a
Reissner-Nordstr\"om solution with $r_H  \simle (ev)^{-1}$, we see that
the Coulomb field has, in a sense, been extended inward too far.
Energetically, it would be preferable to have a core region
extending outward beyond the horizon~[5].  In fact solutions of this
sort, which may be viewed as small black holes lying within
larger magnetic monopoles, can be shown to exist if $v$ is
less than a critical value $v_{cr} \sim M_P$ and if the mass $M$
is not too great~[4].  When they exist, the horizon radius $r_H$ of
these solutions is larger than that of the Reissner-Nordstr\"om
solution with the same value of $M$.  Thus, these solutions
appear to be the natural endpoints to which the instability of
the Reissner-Nordstr\"om solution leads.

       We now turn to the case of multiple magnetic charge.  The
analysis is complicated by the fact that in the $SU(2)$ theory
the only configurations with higher topological charge which are
spherically symmetric (i.e., invariant up to a gauge
transformation under spatial rotations) are the singular
solutions given by Eqs.~(9) and (10)~[6].  There is thus no
spherically symmetric case to which we can restrict our
consideration; instead, we must consider the full perturbation
problem.   This can be done by expanding the action in powers of
the fluctuations about the Reissner-Nordstr\"om solution and
examining the terms quadratic in these fluctuations. (The linear
terms vanish because we are expanding about a solution.)  It is
convenient to use the gauge freedom to require that the orientation
of the scalar field remain the same as in the unperturbed
solution, so that $\delta \Pphi \times \e =0$.
It is also useful to decompose the fluctuation in the gauge field
into parts orthogonal to and parallel to $\e$; thus, we write
$\delta \A_\mu =  \a_\mu + c_\mu \e $
with $\e\cdot \a_\mu =0$.  The fact that $D_\mu \e =0$ (here, and for
the remainder of this discussion, $D_\mu$ is the covariant
derivative defined by the unperturbed vector potential) leads to
the useful result that $\e \cdot D_\mu\a_\nu =0$.

     Several factors simplify the process of extracting the
quadratic terms in the action.   Because $D_\mu\Pphi$,
$\F_{r\mu}$ and $\F_{t\mu}$ all vanish for the unperturbed
solution, terms containing the product of a metric perturbation
and a matter perturbation can only arise from the
$\F_{\theta\phi}\cdot\F^{\theta\phi}$ term in $\L$; it is easy to
see that the only matter field that can enter here is $c_\mu$.
Further, the cross terms between $\a_\mu$ and $c_\nu$, between
$\a_\mu$ and $\delta h$, and between $c_\mu$ and $\delta h$ all
vanish.  The result is that the quadratic part of the action may
be decomposed as
$$ S_{quad}(\a_\mu, c_\nu, \delta \Pphi, \delta g_{\mu\nu})
      = S_1 (c_\mu, \delta g_{\mu\nu}) + S_2( \delta \Pphi) +
         S_3(\a_\mu)
     \eqno(21) $$
Since $c_\mu$ is the component of the fluctuation lying in the
unbroken $U(1)$ subgroup, $S_1$ describes an essentially Abelian
problem; we therefore do not expect it to contain any unstable
modes.  Similarly, $S_2$ is simply the action for a neutral
scalar field in a curved Reissner-Nordstr\"om background, and
easily shown to give no instabilities.

     This leaves us with
$$ \eqalign{
   S_3 &= \int d^4x r^2 \sin\theta \left[
    -{1\over 4}(D_\mu \a_\nu -D_\nu \a_\mu)\cdot
         (D^\mu \a^\nu -D^\nu \a^\mu) \right. \cr
    &\qquad\qquad \left.
     + {1\over 2}e^2v^2\, \a_\mu\cdot \a^\mu
     + e\F_{\theta\phi} \cdot \a^\theta \times \a^\phi  \right]  }
      \eqno(22)$$
where indices are understood to be raised by unperturbed metric
and $\F_{\theta\phi} $ is the unperturbed magnetic field.
Note first
that stability would be manifest if it were not
for the presence of the last term in the integrand.  Indeed,
the instability of the $n=1$ solution sets in as soon as this
driving term can be greater in magnitude than the mass terms for
$\a_\theta$ and $\a_\phi$ just outside the horizon.  With the aid
of the inequality
$$  | e\F_{\theta\phi} \cdot \a^\theta \times \a^\phi | =
  {n \over r^4 \sin\theta} | \e \times \a_\theta \cdot \a_\phi |
   \le  {n \over r^4 \sin\theta}  |\a_\theta| | \a_\phi |
    \eqno(23) $$
it is easily shown that the driving term cannot be dominant, and
thus stability is assured, if $r_H > \sqrt{n}/(ev)$.
Conversely, exponentially growing solutions can be constructed
whenever $r_H < c\sqrt{n}/(ev)$.  An explicit example, which can be
verified by substitution into the field equations derived from
$S_3$, is given by $\a_t=\a_r=0$ and
$$  \eqalign{    \a_\theta
    &= u_n(r,t) \sin^{n-1}\!\theta \,\partial_\theta \e\times\e \cr
     \a_\phi &=  u_n(r,t) \sin^n\!\theta \,\partial_\theta\e }
    \eqno(24) $$
where $u_n(r,t)$ satisfies Eq.~(18), but with $e^2v^2$ replaced by
$ne^2v^2$ in the potential $U(x)$.  As expected, this
solution is not spherically symmetric; under rotation, it
transforms into other linearly independent solutions.   Using
Eq.~(3), we can see that this instability is present whenever $n \simle
(M_P/v)^2$ and
$$    M < M_{inst} = {\sqrt{n}~cM_P^2 \over 2ev}
         + {2\pi n^{3/2} v\over ce}
     \eqno(25)$$

     Some physical understanding of the $n$-dependence of this
result can be obtained by returning to the flat space picture of
a core region of radius $R$ containing nontrivial Higgs and
charged boson fields, with only a Coulomb magnetic field
extending beyond the core.  A variational argument shows that the
value of $R$ which minimizes the energy is proportional to
$\sqrt{n}$.

    This instability has significant implications for the
evolution of a magnetically charged black hole.  A
Reissner-Nordstr\"om black hole will lose mass through the
emission of Hawking radiation~[7].   In the absence of the
classical instability, this process would eventually turn off as
$M$ approached $M_{crit}$, where the Hawking temperature
$$   T_H = {M_P^2\over 2\pi } {\sqrt{M^2 -M^2_{crit}} \over
         \left( M + \sqrt{M^2 -M^2_{crit}} \right)^2 }
     \eqno(26)$$
vanishes, unless it had lost its magnetic charge in the meantime.
Such a discharge could be accomplished by the production of
monopole-antimonopole pairs in the strong magnetic field outside
the horizon, with one particle falling into the hole and the
other moving out to spatial  infinity~[8].  Pair production of
monopoles with magnetic charge $1/e$ becomes significant only in
magnetic fields of magnitude $eM^2_{mon}$~[9].  The field at the
horizon of a hole with charge $n/e$ is this large only if $
M\simle M_{pair} \sim \sqrt{n}M^2_P/v$~[10]. Since this is a factor of
$e$ smaller than $M_{inst}$, pair production is significant only
for black holes which are already classically unstable~[11].

    The classical instability changes this scenario.  Consider
first the case of a hole with unit magnetic charge. Thus, suppose
that a single magnetic monopole falls into a large neutral black
hole, which eventually settles down to a Reissner-Nordstr\"om
solution.  The hole begins to lose mass through the Hawking
process. As the mass falls below $M_{inst}$ and the horizon
contracts within the sphere $r = c(ev)^{-1}$, the instability
causes nontrivial matter fields to begin to outside the horizon.
The black hole is now described by a solution of the type found
in Ref.~4.   Its horizon continues to contract, revealing more
and more of a monopole core.  Its temperature, like that of a
Schwarzschild black hole, increases monotonically. While the
question of its ultimate fate cannot be settled within the
semiclassical approximation, the answer will be the same as for a
Schwarzschild black hole. If the latter can in fact evaporate
completely, then so can our black hole.  When it does so, it
leaves behind a monopole identical to the one which had fallen in
long before.

      This picture is modified slightly if $n>1$.  Because the
unstable modes are not spherically symmetric, the matter fields
which emerge when $M$ falls below $M_{inst}$ do not form a  uniform
shell, but are instead localized about isolated points on the
horizon.  A plausible guess is that  as the horizon contracts
these grow into lumps which can eventually break off as unit
monopoles, thus reducing the magnetic charge of the hole.
Eventually only a single charge is left, and the evolution  proceeds
as described above.

     Furthermore, if $M>M_{crit}$ and $n > (M_P/v)^2$, stability
is assured since $r_H > \sqrt{n}/ev$. Thus,
a black hole could be stabilized by
endowing it with a sufficiently large magnetic charge.  However,
this stabilization is not absolute. Pair production, although
strongly suppressed, is not quite forbidden.  Eventually, enough
of the magnetic charge will have been emitted for the monopole
instability to emerge.

     While these results have been obtained in the context of an
$SU(2)$ gauge theory, they clearly can be extended to
other gauge theories containing magnetic monopoles.  In some
theories with two stages of symmetry breaking it is possible to
have more than one variety of monopole; e.g. a heavy singly-
charged monopole and a somewhat lighter (and spatially larger)
doubly-charged one~[12].   In such theories the Reissner-Nordstr\"om
solutions of higher charge presumably become unstable when their
horizon is comparable to the the size of the lighter monopole,
with the singly charged solution remaining stable until it has
shrunk to the size of the heavier one.

      One can also obtain magnetic monopole solutions in
Kaluza-Klein models~[13].  The question of whether these lead to
similar instabilities is an interesting one, but is beyond the
scope of this letter.

      Thus, the effect we have found leads to a remarkable new
possibility for the evaporation of a black hole carrying a
conserved magnetic charge.  Previously, it had seemed that if
such a hole did not somehow lose its charge the Hawking process
would terminate before complete evaporation was achieved.  We see
now that charge conservation need not be a barrier to complete
evaporation, and that it is quite possible that a magnetically
charged black hole could evaporate completely, leaving in its
place a nonsingular magnetic monopole.

We thank Hai Ren for pointing out an error in a previous version of this
paper.

\centerline{REFERENCES}

\item{1.} G. 't Hooft, \np{79}{276}74; A.M. Polyakov, Pis'ma Zh.
Eksp. Teor. Fiz. {\bf 20}, 430 (1974) [JETP Lett. {\bf 20}, 194
(1974)].

\item{2.} F.A. Bais and R.J. Russell, \pr{11}{2692}75;
 Y.M. Cho and P.G.O. Freund, \pr{12}{1588}75.

\item{3.} J.A. Frieman and C.T. Hill, SLAC preprint SLAC-PUB-4283 (1987).

\item{4.} K. Lee, V.P. Nair and E.J. Weinberg, Columbia and
Fermilab preprint CU-TP-539/FERMILAB-Pub-91/312-A\&T (1991).

\item{5.} It is curious, however, that while the radius and the
precise form of the monopole core vary somewhat with the
couplings in the Higgs potential, the value of $r_H$ at which
instability occurs does not.

\item{6.} E.J. Weinberg and A.H. Guth, \pr{14}{1660}76.

\item{7.} S.W. Hawking, Commun. Math. Phys. {\bf 43}, 199 (1975).

\item{8.} The discharge of electrically charged black holes is
discussed in G.W. Gibbons, Commun. Math. Phys. {\bf 44}, 245 (1975).

\item{9.} I.K. Affleck and N.S. Manton, \np{194}{38}82

\item{10.} W.A. Hiscock, \prl{50}{1734}83.

\item{11.} In conventional units, with $\hbar$ not set equal to
unity, $M_{pair}/M_{inst}$ is of order $\hbar^{1/2}$, reflecting
the quantum mechanical nature of pair production as opposed
to the classical nature of the instability.

\item{12.} G. Lazarides and Q. Shafi, \pl{94}{149}80; G. Lazarides,
M. Magg, and Q. Shafi, \pl{97}{87}80.

\item{13.} D.J. Gross and M.J. Perry, \np{226}{29}83; R.D. Sorkin,
\prl{51}{87}83.

\end